# Digital Sand: The Becoming of Digital Representations

Thomas Østerlie and Eric Monteiro

**Abstract**. The versatility of digital technologies relies on a capacity to represent and subsequently manipulate algorithmically selected physical processes, objects or qualities in a domain. 'Organizationally real' digital representations are those that, beyond the mere capacity to, actually get woven into everyday work practices. Empirically, we draw on a four-year case study of offshore oil and gas production. Our case provides a vivid illustration of Internet of Things (IoT) based visualizations and data driven predictions characteristic for efforts of digitally transforming industrial process and manufacturing enterprises. We contribute by identifying and discussing three mechanisms through which digital representations become organizationally real: (i) *noise reduction* (the strategies and heuristics to filter out signal from noise), (ii) *material tethering* (grounding the digital representations to a corresponding physical measurement) and (iii) *triangulating* (in the absence of a direct correspondence, corroborating digital representations relative to other representations).

1. **Introduction**

"Representation", note Burton-Jones and Grange (2013, p. 636, quoting Weber (2003)), "[is] the essence of all information systems". The versatility of digital technologies relies on the capacity to represent and algorithmically manipulate selected physical processes, objects, or qualities within a domain. How closely they represent the physical domain varies from directly mirroring, to resembling, to decoupled. Pressing the capacity of digital representations to decouple as much as possible is important as this "has the greatest potential to change work's historically tight coupling to the physical and, with it, the work relations of people to objects and each other" (Bailey et al. 2012, p. 1486). In other words, the disruptive potential of digital technologies assumes the capacity of digital representations to decouple from, not merely mirror existing work practices (Borgman 1999).

Several decades of empirical studies of digital technologies in organizations, however, demonstrate how technological potential often fails to translate into organizational change in practice (Zuboff 1988, Leonardi 2012) For digital representations to underpin organizational change in practice, they need to be implicated in consequential decisions and actions within work practices. To become 'organizationally real', digital representations, beyond their mere potential/ capacity for decoupling, need to be incorporated into organizational practices; digital representations are not, but may become, organizationally real. To this end, we pose the following research question: *Through which mechanisms and under what circumstances do digital representations become organizationally real?*

The relevance of our analysis is the expanding scope and depth of digitalization with ever-increasing types and volume of digital representations (Agarwal and Dhar 2014, Brynjolfsson and McAffee 2014, Alaimo and Kallinikos 2017). Particularly relevant to our argument is the sensors of the Internet of Things (IoT). A sensor, quite literally, is a vehicle to generate a digital representation inferred from a physical process, object, or quality (Singh et al. 2014, Monteiro and Parmiggiani 2019). With sensors 'hearing', 'smelling', 'seeing', 'tasting', and with 'tactile sensation', the scope of digital representation potentially expands to approximate human sensory and tactile sensing and embodied action.

Characterized by tacit and embodied skills underpinning judgment, flexibility and sensemaking, so-called knowledge-based work practices have so far remained largely immune to efforts of automation, substitution, and replacement (Autor 2015). We study the IoT-based digital representations implicated in the embodied, tactile, knowledge-based work practice tied to visions of the Second Machine Age (Brynjolfsson and McAffee 2014), Industry 4.0 (Schwab 2016) and the Industrial Internet of Things (Gilchrist 2016).

Empirically, we draw on a four-years case study of digitalization of offshore oil and gas production on the Norwegian continental shelf. The hydrocarbons residing in reservoirs within rock formations several kilometers below the seabed are only accessible through an extensive set of sensors that provide real-time and historical data about selected physical properties. We focus on work practices of *sand monitoring*. The presence of sand in an oil and gas production facility is extremely dangerous. Sand erodes production pipelines, tubes and valves that may threaten human life, economic, and/or environmental value. Sand is routinely monitored at the offshore platforms through inspection rounds with physical sample taking and subsequent laboratory testing. We analyze efforts to replace these practices of sand monitoring with onshore, remotely operated work practices relying on a series of IoT-enabled digital representations (sensor measurement, graphs, and predictive simulation algorithms), increasingly decoupled from the physical sand.

Our case, despite the potential for IoT-based digital representations to decouple from physical sand, is of transformational, not disruptive, change. We contribute by identifying and discussing three mechanisms through which digital representations become organizationally real: (i) *noise reduction* (the strategies and heuristics to filter out signal from noise), (ii) *material tethering* (grounding the digital representations to a corresponding physical measurement) and (iii) *triangulating* (in the absence of a direct correspondence, corroborating digital representations relative to other representations). Taken together, our three mechanisms make up a process model for the becoming of digital representations that in part overlap with but, with the triangulating mechanism, extends existing insights (Zuboff 1988, Bailey et al. 2012, Burton-Jones and Grange 2013).

## 2. Conceptualizing digitalization

Digitalization involves the manipulation of digital representations. Digital representations comprise data but, crucially, also their subsequent algorithmic manipulation. Successive rounds of algorithmic manipulation may result in digital representations taking on the form of an 'algorithmic phenomenon' (Orlikowski and Scott 2015) or a "computational rendition of reality" (Kallinikos 2006, p. 59) decoupled from its origin tied to a physical process, object, or quality. Scholars of digitalization, albeit from different angles and with different formulations, provide strikingly similar insights: Yoo et al. (2010) identify the defining quality of digital technologies their (algorithmic) programmability and layering, Zittrain (2008) characterizes the open-ended extendibility of digital technology by the notion of 'generative', Lusch and Nambisan (2015, p. 160) identify the defining ability of 'liquefaction' of digital representation decoupled "from its physical device or form" (see also Monteiro and Parmiggiani 2019) while Borgman (1999, p. 1) notes the ability of digital representations to "illuminate, transform, or displace reality…[hence] disclose what is distant in space and remote in time".

The above outlined *theoretical* interest into characterizing digitalization is radically boosted by the *empirical* emergence of big data together with data-driven, machine learning-based forms of algorithmic manipulation. IS researchers, to further our understanding, need to *combine* a theoretical grasp of digitalization with an empirical grounding in organizational dynamics, a combination largely missing when it comes to data-driven algorithmic approaches literature reviews consistently find (Günther et al. 2017, Sivarajah et al. 2017, see also Saar-Tsechansky 2015). One helpful manner to assess the representational aspects in practice is to differentiate between situations where the relationship of digital representations to their physical referent are, respectively, indices (i.e. mirroring), icons (i.e. resembling) or symbols (i.e. decoupled) (Bailey et al. 2012).

Historically, digital representations have mirrored the physical domain closely i.e. been indices (Bailey et al. 2012). This mirroring relationship underpinned efforts of digitalization by substituting designated tasks in the physical domain with their directly corresponding digital representation, an approach conceptualized as 'computerization' (Nora 1980, Kling 1996). Computerization was historically tied to their potential to automate a wide set of work tasks (Friedman and Cornford 1989). Braverman (1974) argued in an influential study that the scope of computerization would imply wide-spread deskilling of work tasks. The defining assumption, automation by substituting for manual work task, met with growing critique of both empirical and theoretical nature.

Empirically, scholars demonstrated that the results of computerization were significantly more varied than what Braverman maintained (Noble 1984). Barley's (1986) influential study, for instance, showed how the introduction of similar CT scanners in different hospitals led to different work routines and roles for radiologists. Similarly, the coining of the so-called 'productivity paradox' (Brynjolfsson 1993) underscored the variations in outcomes of computerization: studies found negative, none, and positive correlation between investments in computers and productivity (Kling 1996). A series of studies on computerization demonstrated that digital technologies also involved local appropriation hence was not merely automation (e.g. Gasser 1986, DeSanctis and Poole 1994, Orlikowski 1996).

Digitalization *qua* computerization failed to take on board the transformation, not merely substitution, of work practices, suggesting that it is more useful to conceptualize digital representations as icons resembling but not directly mirroring the physical domain. Early and influentially, Zuboff (1988) argued for the transformational potential of digital technologies as a supplement to their ability to automate. In Zuboff's formulation, digital technologies were different because, beyond automation, they had the potential to 'informate'. They had, Burton-Jones (2014) notes, a capacity to represent that decoupled from the physical. Digital representation, Zuboff's notion of informate underscores, is available and amendable to open-ended algorithmic manipulation. When we input Google search terms, beyond the transaction of returning results from the search, the input may be, and indeed regularly is, used to generate traces of search behavior subject to, for instance, advertisement or surveillance (Orlikowski and Scott 2015, Zuboff 2019).

With icons, digital representations start to decouple from the physical domain (i.e. liquefaction, Lusch and Nambisan 2015). For digital representations qua icons to be organizationally useful, they need to "*faithfully* represent some domain, because they provide a more informed basis for action than unfaithful representations do" (Burton-Jones and Grange 2013, p. 636). So what makes digital representations faithful? A vital organizational condition for digital representations *qua* icons, Leonardi (2012, pp. 12-14) argues, is that they enjoy the immediate recognition through their "similarity with the [physical] object" as "seeing is believing". Similarly, (Bailey et al. 2012, p. 1500) underscore that a necessary organizational condition for digital representations *qua* icons is the possibility of "validating their results against physical objects, people, and their associated processes" hence a "tight coupling between [digital] representations and their referents".

Digital representations *qua* symbols press the decoupling from the physical domain to the limit. Symbols neither mirror nor resemble their physical referents; they are algorithmic phenomenon such as simulations, rankings predictions and models (Agarwal and Dhar 2014, Sugimoto et al. 2016, Alaimo and Kallinikos 2017). The challenge for symbols is to be incorporated organizational practice i.e. become 'organizationally real'. Several scholars, drawing on simulations as empirical instances of symbols, are skeptical to ever arriving at organizational real digital representations qua symbols. Turkle's (2009) work emphasizes the dangers of simulation-based renditions of reality with their strong, seductive capabilities. As users are gradually immersed in simulations, "[f]amiliarity with the behavior of [digital representations] can grow into something akin to trusting them, a new kind of witnessing" (ibid., p. 63). Zuboff's (1988) early work studies analyzed in detail the representational capacity of digital technologies[1]. In her empirical study of digital transformation of pulp factories from experience-based, embodied, tactile handcraft – smelling, tasting, and feeling the temperature of the pulp – into a remotely operated, digitally enabled control room, she notes the unease stemming from "digital [representations] replacing a concrete reality" (p. 63), how digital representations "replace the sense of hands-on" (p. 65) and seeking to "invent ways to conquer the felt distance of the referential function [i.e. the decoupling of the digital representation from the physical referent]". A lack of sensory feedback undermines the expertise and knowledge from physical, hands-on interaction with the technology.

---

[1] The abstraction and manipulation of digital representations are closely linked to notions of the virtual and virtualization. Whereas virtual/ization has many uses, for instance, when physical mechanisms or processes are conducted by computers rather than physically (Overby 2008) or where face-to-face communication is mediated by computers (Jarvenpaa et al. 1998), Bailey et al. (2012) provide a definition where they identify digitization as the creation of computer-based representations of physical phenomena, and a necessary precursor to and hence different from virtuality, i.e. the engagement with these representations. This useful clarification corresponds to Yoo et al.'s (2010) distinction between digitization as the coding into digital formats and digitalization as the processes of engagement made possible by digitization. Our use of digitalization broadly follows Bailey et al.'s (2012) notion of the virtual.

As Burton-Jones (2014, p. 92) observes, "[o]nly recently have studies begun to consider both the representational aspects of [digitalization] and their use in practice." The representational aspects of digitalization remain mere theoretical or technological speculation without an accompanying empirical analysis of the circumstances under which they become consequential for organizational practices (i.e. become 'organizationally real', or, in short, the becoming of digital representations). Our case analyses efforts to replace the practices of physical sample taking of sand monitoring with digital representations, increasingly decoupled from the physical processes. It thus provides a vivid opportunity to analyze central preconditions implicated in visions of The Second Machine Age and Industry 4.0 (Brynjolfsson and McAffee 2014, Schwab 2016).

3. **Research setting and methods**

The empirical focus of this paper is on evolving work practices for monitoring *sand*. Entering offshore production systems through wells drilled thousands of meters beneath the North Sea's seabed, sand particles are swept into the well and along kilometers of pipelines by the fluids entering the well, all the way to the topside processing plant where the sand settles as black, viscous deposits in pipeline bends and the tanks that separates crude oil and natural gas from the other constituents of the fluids. Sand deposits threatens to reduce the offshore plant's processing capacity and oil quality, but more importantly: sand particles rushing at high speeds through the pipelines erodes the piping, eating away at the valves controlling the fluid flow as well as the valve casings hence pose significant risk to human life, the environment as well as business value.

Sand inspections routines have always been part of the offshore roughnecks' daily inspection rounds of the offshore production facility. As production spread to new fields in the 1990s, however, sand in the production system became a more frequent and prominent problem. Traditional sand monitoring routines struggled to keep up as they were time- and resource-demanding. Traditionally, when roughnecks discovered sand in the production equipment, offshore laboratory assistants would be set to a regime of inspecting and emptying the offshore plant's sand traps – cups mounted underneath each flow line where the heavier sand particles settles as fluids rush past – to locate the originating well. It would take time for sand deposits to accumulate in the sand traps, though, so laboratory assistants could only inspect the cups once every eight-hour shift. It would take days or even weeks before the sanding well could be back in production without new sand entering it. In the fiercely competitive petroleum industry of the 1990s, such time-consuming and labour-intensive routines were targeted for reasons of efficiency, quality, and safety. Petroleum companies looked towards digitalization of sand monitoring and mitigation to address the problem.

Digitalization of sand monitoring and mitigating broadly followed the general digitalization trend in the North Sea region. Digitalization of offshore petroleum production remained low in this region well into the 1990s. Massive developments of subsurface data communication cables from the late 1990s and onwards increased data transfer speed and capacity between down-hole sensors and the offshore platform as well as between offshore and onshore production facilities. This sparked a proliferation of digital sensor technologies for monitoring key aspects of production. Connected in large sensor networks stretching across the seabed and deep into individual wells on entire oil fields all the way to onshore operations centers, these remote sensor networks have become constituent parts of a larger digital infrastructure the offshore petroleum industry has built up over the past decades. Today, offshore oil on the Norwegian Continental Shelf is a thoroughly instrumented endeavor with real-time, IoT readings from all phases of the operation gradually moving the industry towards a data-driven, industrial IoT-based production facility.

Reorganization and cost-cutting of operations was a central part of digitalization of offshore petroleum production, resulting in significant amount of personnel previously located on offshore installations were moved to onshore operations centers. Video conferencing, e-mail, and instant messaging were used for communication and collaboration between offshore installations and the mainland. With the availability of real-time sensor data and new engineering applications for visualizing and manipulating this data, onshore engineers could also actively participate in monitoring, diagnosing, and controlling offshore processes. So too, with digitalization of sand monitoring, which can be said to have proceeded along three phases of development outlined in Table 1.

[INSERT TABLE 1 AROUND HERE]

We draw on data from all three periods in analyzing the mechanisms and conditions under which digital sand becomes organizationally real.

**3.1. Data collection**

This paper draws on both authors' sustained engagement with digitalization of the petroleum industry over the past 10 and 25 years respectively. The empirical material reported is predominantly based on the first author's longitudinal case study (January 2009 - December 2013) of digitalization of offshore petroleum production. The study was conducted within Alpha Petroleum Company[2] (APC). APC is a global exploration and production company with 20,000 employees distributed across 35 countries worldwide. APC's main operational base is offshore petroleum production in the North Sea region. The first author conducted fieldwork across three different sites during the four years of the case study (Table 2 summarizes the fieldwork).

[INSERT TABLE 2 AROUND HERE]

*Site 1: APC R&D division headquarters*. The first author conducted fieldwork at APC's R&D division headquarters throughout the entire four years, where he worked closely with researchers and engineers in a department central to digitalization of the corporation's offshore operations. During this period, he went from being an outside observer, to becoming a peripheral participant in one particular research group, before ending up as an outsider again after this research group disintegrated as part of internal reorganizations in the corporation. Throughout the period, however, he retained full access to the R&D division's headquarters, a desk and workstation there, as well as to the people working there.

---

[2] Like all personal names mentioned in this paper, APC is a pseudonym used for issues of confidentiality.

*Site 2: Industry R&D project.* Most of the reported research was funded through a joint industry R&D project, where the first author participated from 2009-2012 along with some of the R&D engineers he was conducting fieldwork among. The R&D project's aim was to develop the next generation of digital technologies for offshore petroleum production, with participants from multiple vendors and petroleum companies central to digitalization of the petroleum industry. Participation in project meetings offered ample opportunity to learn more about digitalization of the offshore petroleum industry. The first author's role in the industrial R&D project was to inform the design of the digital technologies by reporting on technology use in operational settings. To this end, he conducted fieldwork of practical use of digital sensor technologies in operational setting.

*Site 3: Onshore operations center.* During 11 months in March 2009 – February 2010 the first author had a permanent seat among a group of production engineers in one of APC's onshore operations centers. The author followed the engineers' everyday activities, joining them in meetings and more informally during lunch and coffee breaks. With full access to the entire operations center, he was able to speak with engineers from all departments housed at the facility to follow up on events and incidents.

Field notes (Emerson et al. 1995) are the main data material collected during the study. The first author took daily field notes throughout the period in the onshore operations center, and in connection with most meetings and workshops in the R&D project. Initially taking extensive field notes while at the R&D center, note taking ceased as the first author became increasingly integrated with the research team. Most of the field notes were written out at the end of the day, but some remain as jottings in one of many fieldworker's pads. Sometimes, particularly during the fieldwork at the onshore operations center, the first author would tape field conversations believed to be especially relevant to the case study. Most of these were written up as part of the field notes, but conversations regarded as particularly important were transcribed. Towards the end of the case study, we conducted a total of 24 formal semi-structured interviews (Holstein and Gubrium 1995, Kvale and Brinkman 2009) with central stakeholders in APC but also vendor companies and institutions. The interviews supplemented the collected material by going in-depth into particular aspects of the digitalization processes uncovered during the fieldwork.

### 3.2. Data analysis

Our data analysis was path-dependently shaped by prior experience (Suddaby 2006), in our case of interpretative methods (Walsham 2006, Charmaz 2014). Throughout the four years of fieldwork, we iterated between periods of empirical immersion and withdrawing to analyze. We thus enjoyed the flexibility of being open to emerging themes as analysis overlapped with data collection (Eisenhardt 1989). Albeit not part of the data collection, the analysis also draws on the second author's sustained and in-depth engagement with digitalization of the oil and gas industry, including APC, over the past 25 years. The increasingly collective nature of the data analysis was compounded by the fact that data collection ended in December 2013. The authors have spent several of the subsequent years analyzing and re-analyzing the data together. Necessarily reductive, we reconstruct our process of data analysis into three stages to reduce the "inherent creative leaps" (Langley 1999, p. 691).

In the first stage, data analysis was broad and open-ended. With a long-standing affinity with the broad school of practice theory (Orlikowski 2000, Nicolini 2012), we collected data on the organization of work, division of labor on- and offshore, work practices, and the use of digital tools. We coded data manually using colors and annotation in developing descriptive codes. We were sensitive to the contested nature of ongoing digitalization efforts as they came packaged with management's drive for rationalization. Cost-cutting very much included shifting tasks and personnel from off- to onshore, a crucial expectation for heavy investments in digitalization. This was hardly frictionless as for instance one of the biggest labor unions for offshore workers claimed it eroded operational safet[3].

In the second stage, we zoomed in on practices of sand monitoring. This was inductively motivated by the many physical and digital representations of sand in the evolving work practices of sand monitoring. We found the operational engineers' taken-for-grantedness of a conflation between digital representations and physical sand fascinating, as much research on digital representations highlight just the danger of failing to acknowledge the gap between the two. Engaging with literature, we saw how many assumed too much of a dichotomy real/ physical vs. virtual/ digital (Boellstorff 2016). This missed out on the seamlessness that emerged empirically in our case. Deductively inspired by Bailey et al. (2012), we analyzed the seductive capacity of simulation-based renderings of reality (Lahsen 2005). Working with clustering our descriptive codes, we developed concepts about the extensive *work* with sanitizing data pertaining to sand, notably practices for differentiating out or reducing noise. For instance, we found the two descriptive codes 'well status aggregation' and 'false positives reduction' served a similar function of data filtering, albeit in different ways. Finding them both to be ways of reducing different forms of what petroleum professionals regarded as 'noise', we clustered them into the broader concept of 'Signals filtering'. Further working with clustering codes, we identified sociomaterial arrangements to frame how, as our concept read, 'materially tether' sand to its digital representation to ensure faithfulness (Burton-Jones and Grange 2013). Material tethering emerged from our observations that both onshore and offshore personnel sought to tether digital sand data with physical samples observed offshore. However, revisiting our descriptive codes, we realized that another form of material tethering was taking place as APC personnel compared simulated valve erosion with actual erosion in replaced equipment. We thus clustered our descriptive codes along the two concepts of 'real-time' and 'post-hoc tethering'.

---

[3] Commenting on the reduction of offshore employees resulting from automation, a union leader argues that they "request more compelling documentation for the consequences for safety", suggesting automation (with down-sizing) leads to "increased risk of accidents" (see https://e24.no/energi/statoil/tillitsvalgte-de-ansatte-er-blitt-overkjoert/23313142 ).

In the third and final stage, which includes the analysis conducted during the revision process of this paper, we engage more explicitly with deductive imports. Here too, however, the trigger was empirical. In many situations, in the absence of a way to 'materially tether' the digital representations directly to the physical phenomenon (i.e. sand), the engineers moved closer towards a semiotic situation of digital representations leaning on other digital representations in several layers (cf. Baudrillard 1994). How, we asked ourselves, can these digital representations become organizational real when they are not rendered 'faithful' in the manner stipulated by for instance Burton-Jones and Grange (2013)? In our analysis, we developed among others the concepts of 'calibration' to capture the iterative and indirect way of acquiring faithfulness.

The resulting interpretative template from our data analysis is given in Table 3. We use the aggregated constructs of the interpretative template as a vehicle to highlight the mechanisms behind *how* digital representations become organizationally real.

[INSERT TABLE 3 AROUND HERE]

**4. Analysis: Making digital sand organizationally real**

We will here elaborate upon three mechanisms through which sand becomes organizational real. Before progressing, however, we will discuss an underlying premise for the entire analysis: that *digital sensor data resonate with phenomena other than those they are intended to represent* (cf. Østerlie et al. 2012, Parmiggiani and Monteiro 2015). Sensor designers as well as consumers of sensor data refer to this as 'noise'. A sensor generates digital data by registering changes triggered through interaction with its immediate physical surroundings. For instance, the acoustic sand sensor is one of two dominant sensor designs used in offshore operations. Mounted at the outside of pipeline bends, such sensors use hypersensitive microphones to register changes in the sound as the well flow hits the bend. This sound, in turn, is used to measure the amount of sand in the well flow. The electro-resistance sensor is the other dominant sand sensor design used in offshore operations. Mounted inside individual wells, electro-resistance sand sensors use uses Ohm's law to measure changes in electrical resistance across a series of metal probes. Sand erodes the metal probes, and cause a change in electrical resistance. This, in turn, is then transformed into a measure of sand content. However, changes in well flow temperature induce similar changes to that of erosion in resistance and will register as sand even though there is no sand in the well flow. Similarly, being designed around a hypersensitive microphone, the acoustic sand sensor registers all sound changes in the well flow as well as the din of the production machinery transplanting throughout the pipeline system. Data generated by both of the sensor types resonate with phenomena other than just that it is intended to represent (i.e. sand).

Sensor data can rarely be taken at face value. It needs to be contextualized in a larger sociotechnical network of software, organizational practices, and specialized expertise. We analyze this in the form of three mechanisms through which digital representations become organizational real. Our focus on *becoming organizationally real* emphasizes process and longitudinality. We analyze development and everyday use, as well as breakdowns that show when more work is required for digital sand as representation to become organizationally real. We present the three mechanisms identifying defining characteristics across all phases outlined in Table 1.

**Noise reducing**

Mechanisms for reducing, if not entirely eliminating, noise are a central aspect of remote sensing. Few industrial companies will adopt digital sensors viewed to generate noisy and consequently unreliable data. Noise therefore needs to be reduced for sensor data to be implicated in consequential decisions and actions. There are two key aspects to noise reducing mechanisms: signals robustness and signals filtering. *Signals robustness* refers to a sensor system's ability to generate digital data with a robust reference/referent relationship to the phenomenon it is intended to represent. Signals robustness is a factor of sensor design and a measuring scale that hold up spatially across multiple settings. The development and adoption of digital sand sensor technologies during the late 1990s and early 2000s illustrates this. Following increased attention to remote operations from the mid-1990s and onwards, research communities and vendor companies explored different possible technologies for measuring sand content in the well flow:

> *"We already had technology for inspecting pipeline integrity. When we saw the tender for a digital sand monitoring system, we asked ourselves if our existing [electro-resistance] technology could also be used to detect sand in the well flow." (Interview excerpt lead software engineer, sand sensor vendor)*

While conceptual proposals for sand sensors existed already in the mid-1980s, oil and gas companies such as APC engaged with multiple vendors to explored different possible sensing technologies. Towards the late 1990s, two dominant designs emerged from these efforts: the acoustic and the electro-resistance sand sensors. Both technologies transform the changes induced in the sensor to a measurement of sand content through an algorithm. Whereas the elector-resistance sensor's algorithm simply transforms changes in electrical resistance to sand content, signals robustness for the acoustic sand sensor required a much more complex algorithm to detect and eliminate extraneous sound from the well flow and the production equipment.

Robust sensor design draws upon a robust measuring scale. In parallel with the R&D projects to identify and develop potential sand sensor technologies, an international standardization organization was tasked with developing a representative and reliable measurement of sand content. Standardizing sand content measurements may seem easier than it proved to be, as a senior engineer with the standardization organization explained during a project workshop:

> *–The specific way of measuring sand depends on a number of factors. For instance, different approaches are influenced by different factors such as pressure. We tried several approaches, but in the end we landed upon the simplest way of measuring* *sand content: that of grains of sand flowing across a sensing probe every second.*

*(Field note excerpt)*

A robust measurement scale holds up across space and time. Through testing and experimentation in their laboratory setting the standardization organization's research engineers found it difficult to uphold this relation under differing conditions. In the end, measuring sand content as the number of grains flowing across a point in space per second is the product of the research engineers prodding and tweaking of the material arrangements to find the most robust relationship between sandy fluids streaming into a well and sand content as measurable characteristic of the well flow. This material setup, in turn, formed basis for technical qualification of the different vendors' digital sand sensor technologies. Before being accepted as technologically mature, the international standardization organization evaluated each sensor design against this setup. Oil and gas companies required such a technical qualification before acquiring a specific type of sensor for an installation.

Technical qualification is a necessary but not sufficient condition for digital sensor data to be adopted in practice (i.e. implicated in consequential decisions and action thus become 'organizationally real'). It is a necessary condition as no industrial organization will acquire a type of sensor that has not passed technically qualification at the end of its design cycle. However, it is not a sufficient condition if the sensor design does not generate representations that hold spatially across settings. Early failures in introducing digital sand monitoring systems in offshore control rooms are illustrative here.

While control systems to monitor and control offshore processing systems had been digital since the early 1990s, digital sand monitoring expanded the spatial reach of offshore process control beyond the petrochemical processing plant and into the subsurface. The offshore control room is a hectic place. The two operators working there have to remain vigilant at all times to make quick decisions when audio alarms go off as key operating parameters exceed pre-set alarm limits. Alarms go off on a regular basis. APC's chief sand mitigation expert retold these events years later at a workshop:

> *"So you may have sand in the production system." Drawing a jagged, rising line with black marker in a coordinate system on the whiteboard, he continues: "And these are the [sand] data values." Picks up the red marker and draws a red horizontal line across the coordinate system to indicate the alarm level. "And then there is this one single peak above the alarm limit, and then you have triggered an alarm in the offshore control room's process control system. That's just stupid!" Pregnant pause. Looks out at the 10-15 people attending workshop. "What happened, you see, was that they [the offshore control room operators] ignored the alarms, and they said," now speaking with a theatrically exasperated voice, paraphrasing the control room operators, "'The system you have is rubbish [and] we are not able to monitor for sand influx'. So they turned it off, never to use it again." (Fieldnote excerpt, quotes verbatim from tape)*

After only a short period of use, control room operators would turn off the digital sand monitoring system to solely rely on manual routines for inspecting the production equipment again. To qualify for operational use, the sensor vendor had tested and qualified the sensor under laboratory conditions to prove that the referent/reference relation between physical sand and sand data held up. However, the rocks and fluids within a reservoir are by no means as well behaved as a standardization organization's laboratory setup. The digital representations generated by the sensor failed to hold up spatially across the laboratory and production setting despite rigorous testing and technical qualification. Immediately after installing the sensor in a series of wells, APC experienced a number of alarms of sand in the well flow without there being any sand present in the topside production system; a clear indication of a non-robust representational relationship. With further investigation, it turned out that the alarms had been triggered by incorrectly mounted sensors downhole. The sensor vendor redesigned the sensor mounting and developed a set of installation procedures that increased signal robustness.

The control room operators' objections to initial versions of the sand monitoring system also illustrates the second key aspect of nose reducing mechanisms: *signals filtering*. While sensor design seeks to reduce the interference of phenomena other than the one the sensor is intended to register (i.e. noise), it is still rare for sensors to generate completely noise free data. Signals filtering involves mechanisms that weed out irrelevant sensor data. This can be done manually, but with increasing data volumes requires more automated signals filtering to reduce the volume of data that needs to be investigated and sorted out. Despite the early setbacks with introducing digital sand monitoring with the offshore control room, APC pushed forward with their efforts to introduce the system with its onshore production organization where the onshore production engineers and their professional expertise came to be mobilized to help monitor and mitigate sand. Production engineers' daily tasks revolve around planning and prioritizing production to optimally utilize the offshore production plants' processing capacity. Their work is not time critical to the extent of the offshore control room operators, and they have more time to investigate sand alarms. They also have more in-depth and intimate knowledge of individual wells, the particulars of their designs, their production history and all of their idiosyncrasies to better understand the data available:

> *If you only learn one thing from your stay here, a production engineer tells me during lunch break one day, it is that "a well is never simply a well." Well is "only a word". All wells are "different beasts", even though we call them all wells. "It's our job to know all of them." (Senior production engineer, field note excerpt)*

This specialized and intimate knowledge of the material basis of daily production was actualized as their work came to be increasingly focused upon investigating the fluid relation between digital symbols and their reference from the early 2000s and onwards.

Increasing data volumes drive the need for automation to reduce impact on users. So too with signals filtering. While the use of production engineers to sort out and investigate sand alarms were a successful strategy for implementing sand monitoring on many of APC's many offshore installations, the sheer number of wells and alarms on the corporation's massive oilfields threatened to overwhelm the production engineers:

> *"Being able to drill down into the data and to actually correlate different data types is, of course, invaluable. It gives us the chance of actually looking into the data and determine if we need to take action. As long as we monitor erosion on one, two or even a handful of wells, the tool is all we need. But on a field with 120 wells, it's another matter. We need some help to know which wells to pay attention to." (Interview excerpt, senior production engineer)*

To this end, the sand monitoring system vendor developed a frontend system that aggregated all wells into a display traffic lights: all wells in one big dashboard, one traffic light for each well. Instead of having to manually inspect the status of every well, the traffic light showed green for wells with no recent alarms, yellow for wells with a handful of recent alarms, and red for wells with sustained alarms. Similar to the initial sand monitoring system's use of alarm levels to filter away minor amounts of registered sand from sounding an alarm, the traffic light dashboard filtered away peaks above the alarm level and aggregated the results in one joint dashboard.

Signals filtering can under certain circumstances compensate for lacks in signal robustness, for instance when repurposing digital sensor data to bring about new algorithmic phenomena. Oil and gas companies have increasingly turned from real-time monitoring with sensor data, to repurposing sensor data for simulations. Sand monitoring is an example of this.

> *"We used to go about this the wrong way. We just threw down [pipe]lines in the ground, with little or no concern about monitoring their condition. The result is underutilization of the production system." (Field note excerpt, head of Big Ten petroleum company's R&D division)*

Drawing upon the broader concept of predictive maintenance, sand monitoring sought to predict the degree of erosion of individual pipes, bends, and valves to wear out the equipment as much as possible before replacing it. The challenge to such a strategy was, of course, that it was impossible to monitor for wear and tear in real-time. To this end, sand monitoring drew upon digital sand data to simulate the consequence of sand eroding the production equipment. Again, APC's response to what initially amounts to a technical failure is illustrative. Drawing upon diverse fields such as metallurgy and mathematics, a software vendor developed a predictive algorithm that could be connected with the data generated by sand sensors. Yet, working from the assumption of an unequivocal relationship between sand and data, the sand monitoring application quickly faced problems when implemented on an offshore installation:

> *"We quickly realized that input data comes with a lot of uncertainties [read: noise] (...) When the quality of the input data varies, the visualized output is basically meaningless." (Interview excerpt, senior software engineer, sand monitoring system vendor)*

In the spirit of industrial science, APC and the software vendor's engineers did not seek to improve the robustness of sand data. Instead, they relied upon their erosion algorithm as correct and decided to feed the algorithm with an estimate of the amount of sand that would be passing through the production system. Fed with synthetic data, the erosion algorithm proved more robust and reliable. This form of signals filtering resides at the extreme end of noise reduction mechanisms, and completely decouples predicted erosion from actual erosion on the production system. APC coupled this form of noise reducing with another mechanism for making digital sand organizationally real: material tethering.

### 4.1. Material tethering

Material tethering establishes a direct and absolute link between the digital and physical. Whereas noise reducing mechanisms seek robust representations, material tethering is a form of verification targeted at grounding the veracity of digital representations by directly linking it to material manifestations of the same phenomenon. Material tethering is contingent upon access to physical manifestations of the phenomenon in question and unfolds along a real-time/post-hoc continuum depending upon such access. That digital sand monitoring did not replace, but rather came to supplement existing manual sand monitoring practices illustrates the key role played by material tethering in making digital sand organizational real.

While the production engineers' sand monitoring work was centered around digital representations, a notable feature of the transition towards digital sand monitoring was an accumulation of representations, both digital and physical. These representations remained in play at all times, as production engineers also drew upon the topside inspection procedures to verify whether there was sand in the well flow or not. When monitoring for sand there is never a dichotomous separation of the physical and digital. Rather, production engineers move seamlessly between the two. Consider the following snippet from a sand incident during fieldwork at the onshore operations center:

> *–I'm not entirely convinced this is sand, one of the onshore production engineers commented when the sand monitoring software sounded a sand alarm. Continuing:*
> *–Do a flow line test, and initiate an inspection of the sand traps.*

Sand trap inspections were the original sand monitoring mechanisms prior to introducing digital sand monitoring. Sand trap inspections would be triggered if the offshore roughnecks found accumulated sand deposits in the topside processing plant during their daily inspection routine. With no way of telling which well sand these deposits originated from, offshore laboratory assistants would be set to a regime of inspecting and emptying the sand traps mounted where each flow line entered the topside plant. Sand traps were simply cups mounted underneath the flow line where the heavier sand particles would settle as fluids rush past into the processing plant. This inspection regime remained in use together with digital sand monitoring to tether digital sand data to physical manifestations of sand entering the production system. It would, however, take time for sand deposits to accumulate in the sand traps. The offshore laboratory assistant could therefore only inspect the cups once every eight-hour shift. For more immediate feedback on the presence of sand in the well flow (but with no means of telling its source), onshore production engineers would also ask the offshore laboratory assistant to do a well flow sample to verify whether or not there is sand in the well flow.

Real-time material tethering is bidirectional between digital and physical representations. Seated around a big table in the middle of a large operations room, the onshore production engineers would each be working in front of a computer. There were an assortment of material artefacts in the middle of the table that the production engineers would refer to at times; part of a destroyed well screen, printed slide presentations, and a set of vials containing different types of sand. The production engineers would refer to these vials both during sand incidents, and when discussing sand in general. The following vignette drawn from fieldwork at the onshore production center illustrates this:

> *Matt, today's on-call production engineer receives a phone call from the offshore control room. "We have sand deposits in the separator [read: part of the offshore production system]," the control room operator reports. Matt looks puzzled. Looking at the dashboard showing the status of recent sand alarms across the field, he says: "There have been no sand alarms." "But we have sand in the separator", the control room operator insists. Matt is cycling through screens in the sand monitoring application, looking for possible indications of sand, but finding none. Leaning across the table to the set of sand samples, Matt picks one up. The vial's label says 'Silt'. Holding it by its neck, Matt shakes the vial, looking at the quality of the sand swirling within. "What kind of sand is it that you've found?" he asks. "Silt," states the control room operator. "Ah", Matt says, sounding relieved: "Silt is too fine [grained] to register on our sand sensors. There's no erosion danger but let me know when you've located the sanding well so we can take [mitigating] measures." The control room operator confirms – "I'll set the lab[oratory] assistants on it at once" – ending the phone call.*

Post-hoc material tethering is required when physical representations of a phenomenon are not readily available for real-time or near real-time comparison. The predictive sand monitoring strategy illustrates such post-hoc tethering. Predictive sand monitoring was an operational concept that transcended the industry's zero sand tolerance policy making it possible to produce with limited amounts of sand in the well flow. This was a shift away from preventing sand from entering the production system, to managing its consequences. The challenge to this strategy was that it was impossible to monitor for wear and tear on equipment in real time. While the outside of the pipelines and valves were accessible to human inspection – at least for the equipment not located beneath the ground or too deep below the surface – sand erodes the equipment's inner surfaces. There existed sensor equipment that could measure the thickness of the piping, but this required a shutdown of parts of or even the whole production system. It was done as part of bi-annual maintenance shutdowns and was of little help in tethering the algorithm to actual erosion of pipelines or chokes in real-time. Beaming with pride over their achievement, the chief software engineer of the predictive sand monitoring system stated during an interview:

> *"We found that all chokes have eroded as [the system] predicted in 9 out of 9 inspections."*

To make sand erosion present in practice as a reliable phenomenon, production engineers work with a real-time projection that they treat as increasingly unreliable over time, which the sand monitoring software tethers with the actual state of the choke. Here, the digital representation is not sand data, but the simulated erosion of pipes and valves. The material manifestation is not sand accumulations in the production system, but actual erosion on pipes and valves. To verify the erosion algorithm's accuracy, APC would send all chokes from installations with the predictive sand monitoring system to be inspected by experts. These experts would compare actual to predicted erosion as a way of materially tethering the digital representation with the phenomenon after the fact.

### 4.2. Triangulating

Triangulating different digital references may bring out previously undetected or unseen phenomena or aspects of phenomena through indirect and relative links between the digital and the physical. Triangulating follows a logic of corroborating evidence, rather than of reference. The dynamics of the physical phenomena to be investigated underpins and validates the logic of triangulating. Triangulation follows two broad strategies: calibrating and correlating.

Correlating multiple representations indirectly brings out previously unseen physical phenomena or aspects of phenomena. Correlations may be between data points from a single data source, between multiple sources, but may also be between digital representations and theoretical resources. One software vendor's redesign of its original sand monitoring software illustrates this. The vendor's original sand monitoring software represented digital sand as an absolute number, kind of like a digital speedometer. Yet, as its lead software experienced:

> *"The information was presented [in the user interface] in a way they [production engineers] could not relate to. It [the information] was just [presented as] a number, but what does that number mean? They needed to see trends (…)"*
> *(Interview excerpt)*

Visualizing sand as an absolute number made it possible to see whether or not there was sand in the well flow. Yet, correlating data points from a single sand sensor to plot a time series graph enabled the production engineers to see how the amount of sand in the well flow developed over time. Correlating single data points in a graph might have been a simple engineering trick for the software vendor, but it opened a window of opportunity for the production engineers to see new and previously unobserved phenomena. Correlating specific shapes in trended sand data with geo-mechanical theory allowed the onshore production engineers to differentiate between multiple causes for sand entering the well. A steep incline in trended data, for instance, corresponded to an avalanche where the reservoir had collapsed around the well. Repeated spikes of sand data against a background of otherwise low sand influx corresponded to another explanation ('slugging' in this case), and so on.

By correlating trended data with geo-mechanical theory, production engineers could better differentiate the causes of sand influx. Reducing well flow velocity in order to limit fluid drag within the reservoir had previously been the only mitigation strategy. Different causes for sand influx came with their own mitigation strategy to reduce the sand incident's impact on daily production volumes. Whereas reducing well flow velocity would still be used to handle slugging, sand avalanches would instead be mitigated by increasing the well flow velocity:

> *"I'm quite certain we have sand entering the well, but then I look at the down-hole pressure here. [Using the redesigned software, she points at a green trend line plotted in the same coordinate system] I realize that almost no fluids are streaming through the well. I would normally ask the control room operators to choke down [that is: reduce the flow rate on the well] to prevent sand from damaging the production equipment. In this case, however, I am asking them to choke up. We are dangerously close to a shut-in pressure where sand will simply flow back down the pipeline."* (Interview excerpt, production engineer)

The dynamic of the underlying phenomenon (i.e. the sand loaded well flow) validated the correlation of digital representations from multiple sources (trended sand sensor data and pressure sensor mounted within the same well). The sand avalanche was caused by a collapse in the reservoir surrounding the well. Causing sudden and rapid buildup of sand in the well flow threatened to fill the pipelines with sand, as the column becomes too heavy to be lifted out of the well. The greater the well flow's sand content, the more pressure is needed downhole to push the heavy fluid column through the pipelines.[4] Reducing the production rate would limit fluid drag within the reservoir, and hopefully also limit the amount of sand swept along with the fluids being drained out of it. However, reducing well flow velocity would provide too little pressure to lift the sand loaded column towards the surface, causing the sand to flow back into and fill up the pipeline.

---

[4] Contrary to popular belief, hydrocarbons are not pumped out of subsea reservoirs. Rather, the difference between the immense pressure generated within the reservoir by thousands of meters of overburden and that of one atmosphere on the topside platform that pushes fluids out of the reservoir and through the pipelines towards the topside platform.

> *–Looking back at the data collected by the sand sensor system, the data was clear for those of us with knowledge of how sand producing wells behave. Those responsible for mitigating sand often lacked this knowledge. They did not recognize the indicators before the pipeline was filled with sand and irreparable damage had been done. (Field note excerpt, conversation with APC's sand mitigation expert)*

Some correlations were standardized by design, such as the visualizing trended sand data with pressure data in the same plot. Other correlations were standardized by convention. Production engineers learned by on-site training what underlying geo-mechanical phenomena different shapes in trended data correlated with. While many correlations were standardized, correlations were also the go to mechanisms for handling emergent and puzzling situations. Production engineers' depth and breadth of practical as well as theoretical knowledge about the plant and the subsurface was key to such emergent correlations. How the onshore production engineers assigned to one of APC's offshore fields were able to detect and prevent a near disaster illustrates this:

> *Production engineer: "By the time we realized the gravity of the situation (…) there were three and a half millimeters left [of the choke casing]."*
>
> *Researcher: "What would have happened if it [the well flow] eroded through?"*
>
> *Production engineer: "Then you would have a gas leak."*
>
> *Researcher: "Topside?"*
>
> *Production engineer: "Yes, so that would have been a critical situation. It would be like 'every man in the boats', and danger of fire… really not a good situation."*
>
> *(Interview excerpt)*

The offshore plant' test separator had had been filled up with concrete in connection with a failed attempt to abandon a well half a year earlier. Without a test separator, there was no way for to do step-rate testing to calibrate erosion monitoring. Since there had been no major sand incidents, the plant had been run with trace amounts of sand in the well flow without calibration. Yet, the onshore production engineers got an indication of something when the emission data showed too high concentrations of pollutants in the plant's spill water. Correlating spill water data with the plant's choke settings indicated that the wells were producing higher volumes than the choke settings should allow. The production engineers based this inference on the knowledge that excess petroleum in the separators will contaminate the spill water. Suspecting poorly calibrated chokes to be the problem, the offshore control room begins a regime of systematically reducing choke settings to identify the well that was producing too high volumes. The control room identified the well in question on a Friday. On the following Monday, the production engineers check the incoming data from the choking down, and immediately see that something is amiss:

> *"This plot shows the pressure drop, and a pressure drop is an indication that something is going on. The problem here, though, is that the subsurface pressure changes. This shouldn't happen. So I find another similar choking in for that well six months back to compare this with, and do not find the same. That substantiates my interpretation that something is wrong." (Interview excerpt, production engineer)*

Choking the well down actually created a dangerous situation as the eroded choke created a high-pressure beam that quickly bore through the pipe casing. Realizing the gravity of the situation, the production engineers ordered the well shut in and its choke casing inspected: "Three and a half millimeters left (…) mere hours away from catastrophe."

The situation described above was unique. However, emergent correlations tend to stabilize and become an integral part of a community of practice's stock correlations if situations or problems recur repeatedly over time. Investigating a series of false alarms in early stages of digital sand monitoring, the sensor vendor suspected that well flow temperature increases would register as erosion as their sand sensors were built around Ohm's law. By correlating sand alarms with data from temperature sensors along the flow line, the production engineers confirmed this suspicion. This correlation quickly turned into a convention among the production engineers. However, as temperature was plotted in a separate application, in redesigning the sand monitoring software the vendor company standardized the correlation by also plotting temperature in the same visualization as trended sand data.

## 5. Discussion

Touting the 'disruptive' consequences of digitalization smacks of technological determinism insofar as it abstracts from the necessary enabling social, organizational, and technological conditions. In our case, the set of alternative digital representations – sand strobe, graph plots, predictive algorithms – filling the role of sand in everyday sand monitoring did not result in disruptive changes with existing routines based on physical sand specimen sampling. Our digital representations, despite their theoretical capacity, were never dichotomously separated from their physical origin. Rather, the mechanisms through which digital representations come to be implicated in organizational action transcends characteristics of the referent/ reference relationship. Indeed, while the three stages sand monitoring technologies progresses through (see Table 1) broadly follows Bailey et al.'s (2010) progression through three degrees of increasing decoupling of reference/referent, the three mechanisms noise reducing, material tethering and triangulating outlined in the analysis appear to a variable degree across the three phases; i.e. independently of degree of decoupling of referent/reference relationship.

Rather than an ideologically poised dichotomy digital/ physical, the challenge, as Boellstorff (2016, p. 388) points out, is to "understand precisely how the digital can be real". The identification of three mechanisms through which digital representations become organizational real is part of such an understanding. However, we supplement them with a view on the varying conditions under which they become real (or not). By following the development of digital sand monitoring across repeated iterations of development and use, our analysis show that digital representations are at no point organizational *de facto* real. Rather, the representations held to be real repeatedly failed to hold between contexts. For instance, while initial versions of the sand monitoring technology produced robust representations in the development laboratory, it failed to hold under the conditions of offshore control room operators. Yet, with minor modifications to the sensors, the same technology inserted in the daily practices of production engineers came to be implicated in organizational action. Similarly, predictive erosion models, while stable in the development laboratory, failed to hold up operationally. This time around, though, the technology was redeveloped to cater for the vagaries of operational conditions. Yet, it also required work practice change.

By emphasizing the processual character (i.e. becoming) of making digital representations organizational real, we not only recourse to static mechanisms. We also emphasize how digital representations over time come to be implicated in organizational action through various configurations of mechanisms under differing conditions. Such a dynamic perspective leads away from assumptions of representativity as static characteristic that alone determine the adoption of digital representations, as for instance implied through notions such as 'faithful representations' (Burton-Jones and Grange 2013).

**5.1. Similar but not the same: the forging of icons**

Digitalization, drawing on Zuboff's (1988), has historically been dominated by 'automation'. Translated into our perspective this amounts to acknowledging that, despite their theoretical capacity, digital representations have more often than not mirrored the physical objects, processes and qualities. The implication, as captured by our mechanism material tethering, is that efforts of pushing the boundary of 'faithful' (Burton-Jones and Grange 2013) digital representations, come with procedures to closely link the digital representation with its corresponding physical referent; the faithfulness of a digital representation is accordingly an acquired quality.

This insight overlaps with several other studies. Leonardi's (2012) work on the attempts to replace physical (and costly) car crashes in safety design in the automotive industry with simulated crashes is one of few longitudinal studies of simulations in organizations. In his study, Leonardi (ibid., p. 14) underscores the sociotechnical conditions for digital representations in the form of simulations becoming 'real' as "there is no guarantee that people will use the new information [from simulation] or that they will believe it". Simulation models are interesting and relevant to our analysis as they theoretically offer the possibility of decoupling (further) the tie between digital representation and its physical referent. However, a consistent theme in Leonardi's (ibid., p. 62) analysis is how the simulation models need ongoing validation with physical car crashes "to verify the accuracy of their simulations by confirming them with physical tests [of car crashes]". As Bailey et al. (2012), drawing on Leonardi's study, point out "this tight coupling in simulation means that people who create representations are highly dependent on physical referents" (p. 1500). Similarly, in their study of Gehry's use of simulation models in architecture, Boland et al. (2007) underscore the close link maintained with a physical model of the building in question. Our mechanism of material tethering spells out how the link digital/ physical is forged. For instance, during the bi-annual maintenance shutdown, the values for the wear on pipes and chokes predicted by the algorithm is compared directly with the physical measurements of the equipment during maintenance.

If material tethering essentially is an operationalization of the becoming of iconic digital representation as discussed by other scholars, our mechanism noise reducing adds nuances. Material tethering risks portraying the forging of the digital/physical link as overly brittle. Trusting the digital/ physical link, as Leonardi (2012) and Bailey et al. (2012) underscore, is an ongoing achievement. Noise reducing adds an element of robustness to the maintenance of this link. In our case, the widely accepted fault-prone IoT measurements resulted in a tolerance to noise and errors. For instance, the sand probe produced numerous false alarms. The presence of false alarms were not in themselves enough to dismiss the digital/ physical link; there was a certain level of tolerance to the noise distorting the faithfulness of the digital representation from the stand strobe.

In addition to nuancing existing work on digital representations, highlighting how representational faithfulness is an ongoing achievement or acquired quality may also shed some light into why certain kinds of knowledge work has so far has proven immune to automation, substitution, and replacement as promised/threatened by visions industrial digitalization (Brynjolfsson and McAffee, 2014; Schwab 2016; Gilchrist 2016). Brynjolfsson and McAffee (2014, p.23), for instance, posits that "the world is at an inflection point where the effect of (…) digital technologies will manifest with 'full force' through automation". Views of automation, and consequently the substitution and replacement of knowledge work, as the end-point of industrial digitalization leads back to our observation that these debates abstract the social, organizational, and technological conditions necessary for digitalization. While production engineers were able to implicate digital sand data in their everyday sand monitoring activities, digital sand data were by and large unfit for certain algorithmic uses. The story about the predictive erosion algorithm illustrates this. Similarly, an internal R&D group with APC were seeking to expand the algorithms used to ensure operational stability within the processing plant to automatic mitigation of sand incidents. This effort stranded just because the algorithm assumed a faithful referent/ reference relation. However, what these researchers failed to acknowledge was that a faithful referent/ reference relation was an acquired quality of the way the offshore processing plant. Applying their process control algorithm to sand mitigation was more than just applying it to another problem. It was a move out of setting physically designed as a controlled system under cybernetic control into a reality that is by no means as well behaved. Our research supplements visions of automation as the end-point of industrial digitalization with empirically grounded insight into how automation is but one of multiple possible venues for industrial digitalization contingent upon a wider array of conditions.

**5.2. Triangulating: relative and indirect**

Our analysis up till this point resonates deeply with earlier insights as made clear above. The becoming of 'organizationally real' digital representations – simulations, models, predictions and, more generally, algorithmic phenomena – requires crafting and ongoing maintenance of the link digital/ physical. Our mechanisms of material tethering and noise reducing detail the way this grounding operate.

With the triangulating mechanism, however, our analysis opens up to a mode of grounding digital representations not pursued with much energy by scholars of digitalization. Digital representations may be indirectly and relative, not only directly and in absolute terms, tied to the physical referent. In the absence of the direct link grounding the digital representation to the physical through tethering, triangulating demonstrates how the grounding is relative to a number of other digital representations, not only and directly the physical referent. The central dilemma, in the absence of a direct method of material tethering, is discussed by Chang (2004) in his historic analysis of tethering a physical phenomenon (boiling water) to a representation (a measurement with a thermometer). How did one, Chang asks, fix this tethering given rivalling identification of boiling water (e.g. the presence of the first, small bubble vs. the 'explosion' of a big bubble vs. the flow of many smaller bubbles) and alternative designs for thermometers with associated methods of measurement with different measuring characteristics. Fixing the temperature of boiling water directly to a thermometer measurement of 100 degrees centigrade was the end result of these historic experiments, not an available result during the process. The problem, Chang (2004, p. 40) explains, is how to break out of an infinite regress: "But how can one find that first fixed point?... We would like to lay the foundations of a building, but there is no firm ground to out it in. … Are we stuck with an infinite ingress in which one…is validated by another, that one is validated by yet another, and so forth?".

The key to break away from the threat of an infinite regress is what Chang (ibid, p. 45) coins 'epistemic iteration'. Epistemic iteration requires *relative ordering of measurements, not absolute.* In our case, the attempt to measure sand in the well flow directly with a sand strobe corresponds to the problem of fixing the temperature of boiling water to a particular measurement. Fixing the presence of sand directly to a particular sensor measurement - to, in our vocabulary, materially tether, or granting it status as an icon (Bailey et al. 2012) or a faithful representation (Burton-Jones and Grange 2013) - was unattainable, with engineers never trusting the sensor measurement in isolation. Instead of dismissing the sensor measurement, however, it was subject to an ordering where plotted graphs of series of measurements were trusted more than singular data points, as the produced graphs enabled tie-in with trends identified in geo-mechanical models and theory hence were trusted. The singular sand sensor measurement, not to be trusted, was collaborated *relative* to time series of measurements and geo-mechanical models. Our mechanism of triangulating, then, goes beyond the direct and absolute ways of material tethering to accommodate digital representations to become organizational real by an indirect, relative route. Instead of ending up as a mere symbol (Bailey et al. 2012), the singular sand sensor measurement through triangulating was delegated a role in the emerging IoT-based monitoring regime.

**5.3. Revisiting 'disruptive' digitalization**

As pointed out earlier, the disruptive potential of digitalization assumes the decoupling of digital representation from underlying physical processes, qualities, or objects. Our triangulating mechanism has the potential to significantly extend the scope and reach of digitalization. It expands, potentially significantly, the number and types of digital representations becoming organizational real, getting at 'algorithmic phenomena' (Orlikowski and Scott 2015; Kallinikos 2006). Requiring digital representations to be validated directly through procedures of what we dub material tethering significantly limits what digital representations may become 'real', a limitation partially lifted by triangulating. The triangulating mechanism makes inroads into the terrain left out of the dichotomy real vs digital (cf. Boellstorff 2016). It provides an evolutionary trajectory through which new digital representations may become real cumulatively relying on more trusted representations.

In our case, triangulating's capacity for cumulative, evolutionary becoming of digital representations into organizational real was evident in the gradual delegation of sand monitoring to IoT-based prediction algorithms. This has over a period of several years radically reduced the number of offshore personnel by shifting tasks including sand monitoring to onshore control rooms.

## 6. Conclusion

Visions of digitalization – the coming of the Second Machine Age, Industrial Internet of Things, Industry 4.0 and others – predominately underscore the *potential* of digitalization. There is, however, a paucity of empirical studies in IS critically analyzing how, if at all, digital representations move beyond the capacity to represent (i.e. be symbol) to become organizationally real in the sense of woven into everyday work practices. Our process model in the form of the three mechanisms contributes by spelling this becoming out in some detail.

The nature and organizing of sand monitoring changes as work comes to be increasingly centered on digital representations. We offer an account of the socio-technical processes through which digital technologies become infrastructural to work and organizing, i.e. digitalization (Tilson et al. 2010), that takes the process through which digital representations are implicated in organizational action as point of departure. Building upon Burton-Jones and Grange's (2013) observation that representation is the essence of all information systems, a central implication of this paper is therefore that while discussions about the digital/physical or reference/referent relationship may at the surface seem rather academic, such discussions have real-world consequences for understanding the transformation of work and organizing with pervasive digitalization of contemporary organizations.

| Phase | Focus | Key technologies | Central actors |
| --- | --- | --- | --- |
| Phase 1: Monitoring sand content in well flow (mid-1990s-early 2000s) | Real-time measurement of sand content as characteristic of the well flow to replace manual and time-consuming sand monitoring practices. Operational principle of zero sand tolerance implemented as immediate shut-in of sanding well. | Digital sand sensors (acoustic, electrical resistance) Algorithms transforming acoustic/electrical resistance data into measure of sand content in well flow Simple user interface to sensor measurements | Digital sand mitigation within offshore control room operators' production control practices for minimizing disruptions to offshore processing plant. Onshore production engineers supported control room operators in investigating sand alarms. |
| Phase 2: Monitoring events within reservoir (early 2000s) | Combining real-time measurements with geomechanical theory with geo-mechanical knowledge on causes of sand influx transforms digital sand from characteristic of well flow to an indicator of events | Visualizations of sand content data development over time in trends Inclusion of data points from other sensors (temperature, pressure) | Sand mitigation nominally within offshore control room operators' production control practices. I Sand mitigation handled in practice by onshore |

|  | unfolding within the reservoir. Zero sand tolerance policy implemented through diversification of sand mitigation strategies fitted with the kinds of sand event causing sand influx. | to better identify false alarms Dashboard aggregating alarms across all wells on the oil field | production engineers to limit the impact of sand incidents on optimizing daily production volumes. |
|---|---|---|---|
| Phase 3: Predictive sand monitoring (2005 and onwards) | From zero sand tolerance policy towards predicting the effect of producing with limited amounts of sand in the well flow. | Algorithm for predicting erosion on pipeline bends and valves | Sand monitoring exclusively in the domain of optimizing daily production volumes by coordinating erosion of production equipment with bi-annual maintenance shutdowns of the offshore plant. |

**Table 1: Digital sand monitoring phases**

| Site | Period | Materials |
|---|---|---|
| Ethnographic fieldwork among R&D engineers and researchers in R&D division headquarters | January 2009- December 2013 | - Ethnographic fieldnotes (80 typed pages 1.5 line each with handwritten fieldnotes)<br>- 24 formal semi-structured interviews<br>- Documents (archive of 134 separate documents) |

| Active participation in industrial R&D project through workshops and project meetings | January 2009- April 2012 | • Fieldnotes (103 typed pages 1.5 line spacing; 4 r handwritten fieldnotes) <br> • Project reports and presentations (archive of 26 |
|---|---|---|
| Ethnographic fieldwork among petroleum professionals in an onshore production facility | March 2009- February 2010 | • Ethnographic fieldnotes (183 typed pages 1.5 lin of handwritten fieldnotes in each) <br> • 7 taped and transcribed informal interviews |

**Table 2: Overview of data collection and materials collected**

| Constructs | Description | Concepts | Data |
|---|---|---|---|
| Noise reducing | *Sociomaterial arrangements weeding out irrelevant sensor data, sorting relevant from irrelevant data* | Signals robustness | • Formal technology evalua <br> • Add erosion screen to ele <br> • Establish mounting proce |
| | | Signals filtering | • Traffic light aggregating e overview <br> • Use of synthetic data in si |
| Material tethering | *Tying digital representations to the physical representations of the same phenomena.* | Real-time tethering | • Inspect sand traps for san <br> • Well flow sample analyse |
| | | Post-hoc tethering | • Comparison simulated ero equipment |
| Triangulating | *In the absence of a direct means, adding weight to the digital representation by relating it to a other digital representations* | Calibrating | • Remotely Operated Vesse sand to calibrate acoustic <br> • Step rate testing to calibra erosion based on choke th |
| | | Correlating | • Trending of sand measure correspond with different <br> • Correlate sand measureme see if sand alarm is caused <br> • Correlate well flow veloci measurement to see if san |

**Table 3: Summary of mechanisms through which digital representations become organizational real**